\documentclass[aps,prd,twocolumn,superscriptaddress,amsfonts,amssymb,amsmath,showpacs,10pt]{revtex4-2}
\usepackage{silence}
\usepackage{bm}
\usepackage{accents}
\usepackage{mathrsfs}
\usepackage[T1]{fontenc}
\usepackage[normalem]{ulem}
\usepackage{graphicx}
\usepackage[caption=false]{subfig}
\usepackage{booktabs}
\usepackage{dcolumn}
\usepackage{multirow}
\usepackage{hhline}
\usepackage{placeins}
\usepackage{float}

\usepackage[british]{babel}
\usepackage{xcolor}
\usepackage{orcidlink}
\usepackage{xparse}
\usepackage[bb=boondox]{mathalpha}

\def\nn{\nonumber}

\usepackage{hyperref}

\allowdisplaybreaks[1]

\newcommand{\gdep}[1]{\underaccent{\mathsf{NG}}{\hat{#1}}}
\hypersetup{
    colorlinks=true,
    citecolor=blue,
    linkcolor=blue,
    filecolor=magenta,
    urlcolor=blue
}

\begin{document}
\title{Growth factor in teleparallel Gauss-Bonnet gravity}
\author{Shivam Kumar Mishra \orcidlink{0009-0006-4754-4103}}
\email{shivamkumarmishra.mt@gmail.com}
\affiliation{Department of Mathematics, Birla Institute of Technology and Science-Pilani, Hyderabad Campus, Jawahar Nagar, Kapra Mandal, Medchal District, Telangana 500078, India.}

\author{Jackson Levi Said\orcidlink{0000-0002-7835-4365}}
\email{jackson.said@um.edu.mt}
\affiliation{Institute of Space Sciences and Astronomy, University of Malta, Malta, MSD 2080}
\affiliation{Department of Physics, University of Malta, Malta}

\author{B. Mishra\orcidlink{0000-0001-5527-3565}}
\email{bivu@hyderabad.bits-pilani.ac.in}
\affiliation{Department of Mathematics, Birla Institute of Technology and Science-Pilani, Hyderabad Campus, Jawahar Nagar, Kapra Mandal, Medchal District, Telangana 500078, India.}

\begin{abstract}
Teleparallel gravity offers a competing geometric framework on which to build cosmological models. The Gauss-Bonnet invariant captures key aspects of the underlying geometry that has been shown to be an interesting way to form cosmological models beyond $\Lambda$CDM cosmology. In this work, we explore three competing cosmological models in $F(T,T_G)$ cosmology in the context of their evolution of the growth of structure in the Universe. This is a core test of the viability of any cosmological model. In our work, we show how these models are qualitatively competitive with $\Lambda$CDM cosmology for certain ranges of model parameters. Interestingly, the models can arrive at the same level of growth as $\Lambda$CDM while producing possible deviations at intermediate scales.
\end{abstract}

\maketitle

\section{Introduction} \label{sec:intro}
$\Lambda$CDM cosmology provides a high level of consistency with individual observational survey data to a high level of confidence at all cosmic scales where measurements are possible \cite{CosmoVerseNetwork:2025alb,Clifton:2011jh}. For this scenario, cold dark matter (CDM) governs physics on galactic scales, forming large-scale structures \cite{Peebles:2002gy,Baudis:2016qwx,Bertone:2004pz}, whereas dark energy is realised as a cosmological constant $\Lambda$ \cite{Copeland:2006wr} which drives accelerated expansion in the late Universe. On the other hand, the cosmological constant is well known to have internal consistency problems with $\Lambda$CDM \cite{Weinberg:1988cp}, and CDM remains elusive in particle physics detectors  \cite{Gaitskell:2004gd}. In recent years, the increasingly robust appearance of cosmological tensions and anomalies \cite{CosmoVerseNetwork:2025alb,DiValentino:2020vhf,DiValentino:2020zio,DiValentino:2020srs} has prompted a plethora of potential physics scenarios beyond $\Lambda$CDM \cite{Clifton:2011jh,CANTATA:2021ktz}. These different cosmological settings primarily produce different expansion evolution profiles since the strongest statistical tension appears in the value of the Hubble constant $H_0$. On the other hand, the growth of structures in the early Universe and their connection to present large scale structures also expresses a cosmic tension \cite{Abdalla:2022yfr,DiValentino:2020vvd}. Together with the background evolution of these models, the growth of large scale structures is also a core question for cosmological models beyond $\Lambda$CDM.

To confront these tensions, a wide spectrum of different potential solutions have been suggested in the literature \cite{CANTATA:2021ktz,Capozziello:2002rd,Capozziello:2011et,Clifton:2011jh,Nojiri:2010wj,Nojiri:2017ncd}. The vast majority of these classes of models depend on curvature-based geometry coming from the underlying Riemann geometry. Non-Riemannian geometries have remained poorly studied for many decades due to their lack of a mature geometric base. Different gravitational connections were probed, leading to the exploration of non-Riemannian geometries and their potential use in producing cosmological models. An important development in non-Riemannian geometries has been the development of teleparallel gravity (TG) \cite{Bahamonde:2021gfp,Aldrovandi:2013wha,Cai:2015emx,Krssak:2018ywd}. Over these years, TG has gained immense popularity. TG relies on the exchange of the curvature-based Levi-Civita connection $\bar{\Gamma}^{\sigma}{}_{\mu\nu}$ (over-bars denote quantities based on the Levi-Civita connection in this work) with the torsion-based teleparallel connection $\Gamma^{\sigma}{}_{\mu\nu}$. While Riemannian geometries are better developed, TG may provide a more intuitive geometric framework on which to build cosmological theories.

The TG connection expresses geometric torsion while also being curvature-free and satisfying metricity \cite{Aldrovandi:2013wha}. A particular combination of the TG connection exists that produces the teleparallel equivalent of General Relativity (TEGR), which is dynamically equivalent to general relativity (GR) at the level of the classical field equations \cite{Capozziello:2022zzh}. In this way, both GR and TEGR have identical properties at astrophysical and cosmological scales, with possible differences arising in any future IR completion of either model \cite{Mylova:2022ljr}. TEGR is based on a linear appearance of the torsion scalar $T$. Similar to the way that $f(R)$ models emerge from the generalization of the Einstein-Hilbert action, TEGR can be generalized to $f(T)$ gravity directly \cite{Ferraro:2006jd,Ferraro:2008ey,Bengochea:2008gz,Linder:2010py,Chen:2010va,Basilakos:2013rua,Cai:2015emx,Bahamonde:2019zea,Paliathanasis:2017htk,Farrugia:2020fcu,Bahamonde:2021srr,Bahamonde:2020bbc}. In this scenario, the Einstein-Hilbert action is replaced by an arbitrary function of the torsion scalar $T$. In contrast to $f(R)$ gravity, $f(T)$ gravity produces generically second-order theories of gravity. In recent years, $f(T)$ gravity has been challenged by the possibility of strong coupling \cite{BeltranJimenez:2020fvy,Hu:2023juh,Guzman:2019ozl,Danieli:2025mov} where the degrees of freedom of the underlying theory do not all materialize in the background and linear equations of motion of the system.

There are many intriguing aspects of $f(T)$ gravity and its cosmology \cite{Bahamonde:2021gfp,Cai:2015emx,Krssak:2018ywd}, but structural problems pose a challenge to its possible realization. Within this context, it may also be interesting to consider other contributors to the broader TG framework, such as the teleparallel Gauss-Bonnet scalar invariant $T_G$ \cite{Kofinas:2014daa,Kofinas:2014aka,Kofinas:2014owa}. Analogous to $f(T)$ gravity, an $F(T,T_G)$ class of gravity models can be constructed directly \cite{Bahamonde:2016kba,delaCruz-Dombriz:2017lvj,delaCruz-Dombriz:2018nvt}. The $T_G$ scalar was first derived in Ref.~\cite{Kofinas:2014owa} where the $F(T,T_G)$ equations were also derived. This general class of models was applied to a background cosmological setting in Ref.~\cite{Kofinas:2014aka}, with the Friedmann equations and some example dynamical system analyses. The general class of models was later explored through different reconstruction approaches in Ref.~\cite{delaCruz-Dombriz:2017lvj,delaCruz-Dombriz:2018nvt} where several models were proposed. The perturbative sector has been derived in Ref.~\cite{mishra2025gaugeinvariantperturbationsfttg} where an initial analysis of the stability of $F(T,T_G)$ models was performed. The astrophysical properties of these models was investigated in Ref.~\cite{Bahamonde:2016kba}, while the gravitational waves signature of this class of models was probed in Refs.~\cite{Mishra:2025,Farrugia:2018gyz}.

The growth of large-scale structures is a key prediction of any realistic class of cosmological scenarios. Given the growing body of work in $F(T,T_G)$ cosmology, we explore the question of how realistic physical models compare with $\Lambda$CDM in these classes of models. Matter perturbations couple to scalar perturbations to form large-scale cosmological structures, which evolve over time. This growth of structure can be probed through the growth factor $\mathcal{D}(a)$, which encodes the evolution of linear density matter fluctuations. By considering different leading models of the literature, we explore how these evolution profiles contrast with one another and with the standard cosmological scenario. To do this, we first consider the underlying physics in Sec.~\ref {sec:tech_intro} where TG and its route to $F(T, T_G)$ gravity are described. The background and linear perturbation equations of $F(T,T_G)$ gravity are also presented here for the scalar sector. It is through these equations that the M\'esz\'aros equation can then be derived in Sec.~\ref {sec:meszaros}. This is a core relationship in the evolution of large-scale structures and in how matter perturbations relate to present-day surveys that map the distribution of galaxies in the Universe. In Sec.~\ref{sec:num_results}, three leading cosmological models within the $F(T,T_G)$ gravity literature are explored numerically to assess how they compare with the $\Lambda$CDM evolution of the growth of structures in the Universe, and how different model parameter values affect this evolution. Overall, the growth of structures in these models appears competitive with the standard cosmological model, which may be another indication of the viability of the predictive power of these models. Finally, the main results are summarized in Sec.~\ref{sec:conclusion} where further analysis of these results is discussed.

\section{\texorpdfstring{$F(T,T_G)$}{F(T,TG)} Gravity and Cosmology} \label{sec:tech_intro}
Let us briefly review $F(T, T_G)$ gravity. In Einstein’s general relativity (GR), gravity is encoded in the metric tensor, which determines distances and angles on the spacetime manifold $ \mathcal{M}$. The compatible connection is the torsion-free Levi-Civita connection, whose curvature encapsulates gravitational effects. For each point $ p \in \mathcal{M} $ within a local coordinate chart denoted as $ \{x^{\mu}\} $, the tangent space $ T_{p}\mathcal{M} $ represents a vector space that is associated with the manifold. In teleparallel construction, we use Greek indices $(\mu, \nu, \rho, \dots)$ for holonomic indices corresponding to the spacetime manifold, while Latin indices $(a, b, c, \dots)$ refer to a non-holonomic basis $\{e_{a}\}$ of the tangent space $ T_{p}\mathcal{M} $. This non-holonomic basis is related to the coordinate basis via the tetrad components $e_{a} = {e_{a}}^{\mu} \partial_{\mu}$. In four dimensions, both Latin and Greek indices take values from $0$ to $3$, with $x^0$ representing the time coordinate.

Let $\{e_a(x)\}$ be a non-holonomic basis for the tangent space $ T_pM $ at the point $ x $. The covariant derivative of the basis vector $ e_a $ is defined  as follows \cite{Izumi:2012qj},
\begin{equation}
    \nabla e_a(x) = \omega^b{}_a(x)\, e_b(x)\,,\label{eq_1}
\end{equation}
where $ \omega^b{}_a(x) $ is known as the connection one-form, or spin connection. The components of this connection are given by,
\begin{equation}
    \omega^b{}_a(x) = \langle e^b,\, \nabla e_a \rangle = \omega^b{}_{a\mu}(x)\, dx^\mu,\label{eq_2}
\end{equation}
Here, $\{e^b(x)\}$ is the coframe that is dual to the basis $\{e_b(x)\}$, and $\{dx^\mu\}$ serves as the dual basis to the coordinate basis $\{\partial_\mu\}$. Like the frame $ e_a $, the coframe connects to the dual coordinate basis as $ e^a = e^a\,_\mu\,dx^\mu $.

In teleparallel gravity, the spin connection is purely inertial, meaning it accounts only for inertial effects and has vanishing curvature \cite{Aldrovandi:2013wha},
\begin{equation}
R^{a}\,_{b\alpha\beta} = 2\partial _{[\alpha }\omega ^{a}\,_ {|b|\beta ]} +2 \omega ^{a}\,_{c [\alpha } \, \omega^{c}\,_{|b|\beta ]} = 0\, .
\label{eq_3}
\end{equation}

The space-time metric relates to the Minkowski metric through the tetrads as
\begin{equation}
    g_{\alpha\beta} = \eta_{ab}\, e^a{}_\alpha\, e^b{}_\beta\,,
\label{eq_4}
\end{equation}
which enforces orthonormality of the tetrads. Taking determinants gives
\begin{equation}
    e:=\det(e^{a}{}_{\alpha})=\sqrt{|g|}\,.
\label{eq_5}
\end{equation}

Using tetrads, the teleparallel connection $\Gamma^\rho{}_{\beta\alpha}$ can be related to the spin connection $\omega^a{}_{b\alpha}$ \cite{Aldrovandi:2013wha},
\begin{equation}
    \Gamma^\rho{}_{\beta\alpha} = e_a{}^\rho \partial_\alpha e^a{}_\beta + e_a{}^\rho \omega^a{}_{b\alpha} e^b{}_\beta\,,
\label{eq_6}
\end{equation}

which shows that tetrads are covariantly constant (teleparallel condition), $\nabla_X e_a=0$. Direct computation then shows the teleparallel connection is flat,
\begin{equation}
R^{\lambda}\,_{\delta\alpha\beta} = 2\partial _{[\alpha }\Gamma ^{\lambda}\,_ {|\delta|\beta ]} +2 \Gamma ^{\lambda}\,_{\zeta [\alpha } \, \Gamma ^{\zeta}\,_{|\delta|\beta ]} = 0\,.
\label{eq_7}
\end{equation}

For a given metric in Eq. \eqref{eq_4}, the tetrad is not unique. Covariance requires choosing a suitable spin connection to compensate for the tetrad choice. Naively adopting Minkowski tetrads can introduce spurious inertial effects. In flat space-time, Eq. \eqref{eq_4} is satisfied only when tetrads correspond to local Lorentz transformations \cite{Aldrovandi:2013wha}. The inertial spin connection can be written as
\begin{equation}
    \omega^{a}{}_{b\alpha} := \varLambda^{a}{}_{c} \partial_{\alpha} \varLambda_{b}{}^{c}\,.
\label{eq_8}
\end{equation}
Here $\varLambda^{a}{}_{c}$ denotes Lorentz transformations. Due to flatness \eqref{eq_3}, one can always choose a frame where the spin connection vanishes; tetrads satisfying this are called proper tetrads and define the Weitzenböck gauge.

The teleparallel connection \eqref{eq_6} has torsion,
\begin{equation}
  T^{\lambda}{}_{\alpha\beta} = \Gamma^{\lambda}{}_{\beta\alpha} - \Gamma^{\lambda}{}_{\alpha\beta}\,.
\label{eq_9}
\end{equation}

Using metric compatibility and torsion definitions, it is useful to define the contortion tensor via
\begin{equation}
 \mathcal{K}^{\lambda}{}_{\alpha\beta}= \frac{1}{2}(T_{\alpha}{}^{\lambda}{}_{\beta}+T_{\beta}{}^{\lambda}{}_{\alpha}-T^{\lambda}{}_{\alpha\beta})\,.
\label{eq_10}
\end{equation}
The Ricci scalar derived from the standard Levi-Civita connection is expressed as follows,
\begin{equation}
\bar{R} = -T + 2 \bar{\nabla}_{\alpha} T^{\beta \alpha}{}_{ \beta}\,,
\label{eq_11}
\end{equation}
where the torsion scalar is
\begin{equation}
\begin{aligned}
T &:= \frac{1}{4} T^{\lambda \alpha \beta} T_{\lambda \alpha \beta} + \frac{1}{2} T^{\lambda \alpha \beta} T_{\beta \alpha \lambda} - T_{\alpha}{}^{\alpha \lambda} T^{\beta}{}_{\beta \lambda}\,.
\end{aligned}
\label{eq_12}
\end{equation}

The torsion scalar can also be written using the superpotential $S_\alpha{}^{\beta\lambda}$ \cite{Bahamonde:2021review},
\begin{equation} 
T =  S_{\alpha}{}^{\beta \lambda} T^{\alpha}{}_{\beta \lambda}\,,
\label{eq_13}
\end{equation}
with
\begin{equation} 
S_{\alpha}{}^{\beta \lambda} =\frac{1}{2}( \mathcal{K}^{\beta \lambda}{}_{\alpha} - \delta_{\alpha}{}^{\beta} T_{\sigma}{}^{\sigma \lambda} + \delta_{\alpha}{}^{\lambda} T_{\sigma}{}^{\sigma \beta})\,.
\label{eq_14}
\end{equation}

Equation \eqref{eq_11} implies
\begin{equation}
e\, \bar{R} = -e\,T + 2\partial_{\alpha}(e\,  T^{\beta \alpha}{}_{ \beta})\,.
\label{eq_15}
\end{equation}
Thus, the torsion-scalar action is dynamically equivalent to GR, known as the teleparallel equivalent of general relativity (TEGR). Extending TEGR leads to $F(T)$ gravity, where the Lagrangian is generalized to an arbitrary function $F(T)$. Generalizing this construction, Ref. \cite{Kofinas:2014owa} introduced the teleparallel analogue of the Gauss-Bonnet invariant $\bar{G}=\bar{R}^{2}-4\bar{R}_{\alpha\beta}\bar{R}^{\alpha\beta}+\bar{R}_{\alpha\beta\lambda\rho}\bar{R}^{\alpha\beta\lambda\rho}$ using a torsion scalar $T_G$, satisfying
\begin{equation}
e\bar{G} = eT_{G} + \text{total divergence}\,,
\label{eq_16}
\end{equation}
with explicit form
\begin{eqnarray}
&&\!\!\!\!\!\!\!\!\!
T_G=(\mathcal{K}^{p}_{\,\,\,\,\,\,ea}\mathcal{K}^{eq}_{\,\,\,\,\,\,\,b}
\mathcal{K}^{r}_{\,\,\,\,\,\,fc}\mathcal{K}^{fs}_{\,\,\,\,\,\,\,d}
-2\mathcal{K}^{pq}_{\,\,\,\,\,\,\,a}\mathcal{K}^{r}_{
\,\,\,\,\,\,eb}\mathcal{K}^{e}_{\,\,\,\,\,\,fc}\mathcal{K}^{fs}_{\,\,\,\,\,\,\,d}
\nn\\
&& \ \ \ \ \,+2\mathcal{K}^{pq}_{\,\,\,\,\,\,\,a}\mathcal{K}^{r}_{\,\,\,\,\,\,eb}\mathcal{K}^{es}_{\,\,\,\,\,\,\,f}\mathcal{K}^{f}_{\,\,\,\,\,\,cd}
\nn\\
&& \ \ \ \ \,+2\mathcal{K}^{pq}_{\,\,\,\,\,\,\,a}\mathcal{K}^{r}_{\,\,\,\,\,\,eb}\,\partial_d\,\mathcal{K}^{es}_{\,\,\,\,\,\,\,c})\delta^{\,a\,b\,c\,d}_{p\,q\,r\,s}\,,
\label{eq_17}
\end{eqnarray}
where $\delta^{\,a\,b\,c\,d}_{p\,q\,r\,s}$ is the generalized Kronecker determinant tensor.

Since $T_G$ is topological in four dimensions, linear dependence does not modify dynamics. Hence, we consider the action
\begin{eqnarray}
S = \frac{1}{2\kappa^{2}} \int_{\mathcal{M}} d^{4}x\, e\, F(T,T_G) + S_m\,,
\label{eq_18}
\end{eqnarray}
which generalizes TEGR and is different from both $F(T)$ and curvature-based $F(\bar{R},\bar{G})$ analogues. Here, $\kappa^{2} = 8\pi G$ and $S_m$ is the matter Lagrangian, which defines the matter energy-momentum tensor via
\begin{equation}
\frac{\delta S_m}{\delta e^a{}_{\alpha}} =  e \,e_a{}^\beta\,\Theta_\beta{}^{\alpha}\,=e\,\Theta_a{}^{\alpha}\,.
\label{eq_19}
\end{equation}
Varying the action with respect to tetrads, the field equations are given by \cite{Bahamonde:2016kba, Bahamonde:2021review}
\begin{align}
\mathbb{E}_a{}^{\alpha}
&:= \frac{1}{e}F_{T}\partial_{\beta}(e S_{a}{}^{\alpha\beta})
+ S_{a}{}^{\alpha\beta} (\partial_{\beta}F_{T})
+ F_{T}T^{b}{}_{\beta a}S_{b}{}^{\alpha\beta}
 \nonumber\\
&\quad
+ \frac{1}{4}F\, e_a{}^\alpha- \frac{1}{2}F_{T_G}\,
\delta^{mbcd}_{ijk\ell}\,
e_{d}{}^{\alpha}\,
\mathcal{K}^{ij}{}_{m}\,
\mathcal{K}^{k}{}_{eb}\,
\partial_a \mathcal{K}^{e\ell}{}_{c} \nonumber\\
&\quad
+ \frac{1}{4e}\partial_{\beta}\Big[
\eta_{a\ell}
\big(
\mathcal{Y}^{b[\ell h]} - \mathcal{Y}^{h[\ell b]} + \mathcal{Y}^{\ell[b h]}
\big)
e_{h}{}^{\beta} e_{b}{}^{\alpha}
\Big] \nonumber\\
&\quad
+ \frac{1}{4e} T_{iab} e_{h}{}^{\alpha}
\big(
\mathcal{Y}^{b[i h]} - \mathcal{Y}^{h[i b]} + \mathcal{Y}^{i[b h]}
\big)
=\frac{\kappa^2}{2} \Theta_a{}^{\alpha}\,
\label{eq_20}
\end{align}

with definitions
\begin{align}
\mathcal{Y}^{b}{}_{ij} &= e F_{T_{G}} \mathcal{X}^{b}{}_{ij} - 2 \delta^{cabd}_{elkj} \partial_{\alpha} ( e F_{T_G} e_{d}{}^{\alpha} \mathcal{K}^{el}{}_{c} \mathcal{K}^{k}{}_{ia} ) \,,
\label{eq_21}
\end{align}
and
\begin{align}
&\mathcal{X}^{a}{}_{ij} 
:= \frac{\partial T_G}{\partial \mathcal{K}_a{}^{ij}} \nonumber\\
&\qquad= \mathcal{K}_j{}^{e}{}_{b}\,\mathcal{K}^{k}{}_{fc}\,\mathcal{K}^{fl}{}_{d}\,\delta^{abcd}_{iekl}
+ \mathcal{K}^{e}{}_{ib}\,\mathcal{K}^{k}{}_{fc}\,\mathcal{K}^{fl}{}_{d}\,\delta^{bacd}_{ejkl} \nonumber\\
&\quad
+ \mathcal{K}^{k}{}_{ec}\,\mathcal{K}^{ef}{}_{b}\,\mathcal{K}_{j}{}^{l}{}_{d}\,\delta^{cbad}_{kfil}
+ \mathcal{K}^{f}{}_{ed}\,\mathcal{K}^{el}{}_{b}\,\mathcal{K}^{k}{}_{ic}\,\delta^{dbca}_{flkj} \nonumber\\
&\quad
+ 2 \,\mathcal{K}^{k}{}_{eb}\,\mathcal{K}^{el}{}_{f}\,\mathcal{K}^{f}{}_{cd}\,\delta^{abcd}_{ijkl}
- 2\,\mathcal{K}^{k}{}_{eb}\,\mathcal{K}^{e}{}_{fc}\,\mathcal{K}^{fl}{}_{d}\,\delta^{abcd}_{ijkl} \nonumber\\
&\quad
- 2\,\mathcal{K}^{ke}{}_{b}\,\mathcal{K}^{j}{}_{fc}\,\mathcal{K}^{fl}{}_{d}\,\delta^{bacd}_{keil}
- 2\,\mathcal{K}^{ef}{}_{c}\,\mathcal{K}^{k}{}_{ib}\,\mathcal{K}^{jl}{}_{d}\,\delta^{cbad}_{efkl} \nonumber\\
&\quad
- 2\,\mathcal{K}^{fl}{}_{d}\,\mathcal{K}^{k}{}_{eb}\,\mathcal{K}^{e}{}_{ic}\,\delta^{dbca}_{flkj}
+ 2\,\mathcal{K}^{ke}{}_{b}\,\mathcal{K}_{j}{}^{l}{}_{f}\,\mathcal{K}^{f}{}_{cd}\,\delta^{bacd}_{keil} \nonumber\\
&\quad+ 2\,\mathcal{K}^{k}\,_{eb}\,\partial_{d}\,\mathcal{K}^{el}\,_{c}\,\,\delta^{abcd}_{ijkl}
+ 2\,\mathcal{K}^{ke}\,_{b}\,\partial_{d}\,\mathcal{K}_{j}{}^{l}{}_{c}\,\delta^{bacd}_{keil}\nonumber\\
&\quad
+ 2\,\mathcal{K}^{el}{}_{f}\,\mathcal{K}^{k}{}_{ib}\,\mathcal{K}^{a}{}_{cd}\,\delta^{fbcd}_{elkj}
+ 2\,\mathcal{K}^{fc}{}_{d}\,\mathcal{K}^{d}{}_{eb}\,\mathcal{K}^{el}{}_{i}\,\eta_{mj}\,\delta^{dbma}_{fckl}\,.
\label{eq_22}
\end{align}

Decomposing the field equations into symmetric and antisymmetric parts is vital for modified teleparallel gravity. The antisymmetric components govern frame-dependent effects and are identical to the equations derived from varying with respect to the flat spin connection in covariant formulations \cite{Golovnev:2017}.
\begin{eqnarray}
    \mathbb{E}_{\alpha\beta} = \mathbb{E}_{(\alpha\beta)} + \mathbb{E}_{[\alpha\beta]}\,.
\label{eq_23}
\end{eqnarray}

To maintain the symmetry of the energy-momentum tensor and to restore local Lorentz invariance, it is essential that the following condition is satisfied 
\begin{equation}
     \mathbb{E}_{[\alpha\beta]}=0
\label{eq_24}
\end{equation}
to restore consistency.
\subsection{Background Cosmology}
In what follows, we consider the Friedmann--Lemaître--Robertson--Walker (FLRW) line element describing the expanding Universe 
\begin{equation}
ds^{2} = - dt^{2} + a^{2}(t) \delta_{\hat{i}\hat{j}} dx^{\hat{i}} dx^{\hat{j}}\,,
\label{eq_25}
\end{equation}
where $a(t)$ denotes the cosmological scale factor, and the hatted indices correspond to the three spatial coordinates. The above line element follows from choosing the diagonal tetrad
\begin{equation}
\left[e^{a}\,_{\alpha}\right] = \mathrm{diag}(1, a(t), a(t), a(t))\,,
\label{eq_26}
\end{equation}
using relation \eqref{eq_4}. The corresponding inverse tetrad reads
$\;\left[e_{a}\,^{\alpha}\right] = \mathrm{diag}(1, a^{-1}(t), a^{-1}(t), a^{-1}(t))$, and its determinant is $e = a^{3}(t)$. Inserting the tetrad \eqref{eq_26} into the definitions given in \eqref{eq_12} and \eqref{eq_17}, we find
\begin{eqnarray}
\label{eq_27}
T &=& 6 \frac{\dot{a}^{2}}{a^{2}} = 6 H^{2}\,, \\
\label{eq_28}
T_G &=& 24 \frac{\dot{a}^{2}}{a^{2}} \frac{\ddot{a}}{a} = 24 H^{2} \left(\dot{H} + H^{2}\right)\,,
\end{eqnarray}
where the Hubble parameter is defined as $H = \displaystyle\frac{\dot{a}}{a}$, and an overdot represents differentiation with respect to time $t$.

Assuming that the matter content of the Universe is a perfect fluid with energy density $\rho$ and pressure $p$. i.e. 
\begin{equation}
    \Theta_{\alpha\beta}=(p+\rho)\mathbb{n}_\alpha\,\mathbb{n}_\beta+p\,g_{\alpha\beta}\label{eq_29}
\end{equation}
subject to the condition, $\mathbb{n}^\alpha\,\mathbb{n}_\alpha=-1$. The field equations in the background are given by \cite{Kofinas:2014daa}
\begin{equation}
F - 12 H^{2} F_{T} - T_{G} F_{T_{G}} + 24 H^{3} \dot{F_{T_{G}}} = 2 \kappa^{2} \rho\,,
\label{eq_30}
\end{equation}
\begin{align}
F &- 4(\dot{H} + 3H^2)F_T - 4H\dot{F}_T 
- T_G F_{T_G} + \frac{2}{3H}T_G \dot{F}_{T_G} \nonumber\\
&+ 8H^2 \ddot{F}_{T_G} = -2\kappa^2 p\,
\label{eq_31}
\end{align}

In what follows, we assume that the background field equations hold. Under this assumption, the background matter density and pressure, $\rho$ and $p$, can be interchanged with their corresponding geometric contributions in the background field equations, and vice versa.

\subsection{Inhomogeneous Scalar Perturbations}

Scalar perturbations, which arise from variations in density and pressure, drive the formation of large-scale structures and generate temperature anisotropies in the CMB, offering direct insight into early Universe physics. To study structure growth, we consider the linearly perturbed metric in the Newtonian gauge.

\begin{equation}
\underset{NG}{\left[\delta g_{\alpha\beta}\right]} = \left[\begin{array}{cc}
-2 \gdep{\Phi} & 0\\
0 & 2a^{2}\gdep{\Psi}\delta_{\hat{i}\hat{j}}
\end{array}\right]\,. \label{eq_32}
\end{equation}

To prevent issues with the integrated Sachs-Wolfe effect \cite{Zheng_2011}, the selection of metric perturbation is determined by the choice of tetrad 

\begin{eqnarray}
   \underset{NG}{\left[\delta\, e^{a}\,_{\alpha}\right]}  = \left[\begin{array}{cc}
     \gdep{\Phi} & a\partial_{\hat{i}} \gdep{\beta}  \vspace{1ex}\\
    \delta^{\tilde{l}}{}_{\hat{i}}\partial^{\hat{i}}\gdep{\beta} & a\delta^{\tilde{l}\,\hat{i}}\,\gdep{\Psi}\delta_{\hat{i}\hat{j}}
    \end{array}\right]. \label{eq_33}
\end{eqnarray}

To derive the system’s field equations, we introduce perturbations in the energy–momentum tensor in an analogous manner
\begin{equation}
 \underset{NG}{\left[\delta \Theta_{\alpha\beta}\right]} = \left[\begin{array}{cc}
\gdep{\delta \rho} + 2 \rho  \gdep{\Phi} &  - a (\rho + p) \, \partial_{\hat{i}}\,\gdep{U} \vspace{1ex} \\
 - a (\rho + p) \, \partial_{\hat{i}}\,\gdep{U} & a^{2}\left(\gdep{\delta P} \,  + 2 p \gdep{\Psi} \delta_{\hat{i}\hat{j}}\right)
\end{array}\right]
\label{eq_34}
\end{equation}

Where $\gdep{U}$ comes from the scalar part of perturbations of the 4-vector $\mathbb{n}$. Inserting the above in the  general field equations, we get the on-shell linearized equations as \cite{mishra2025gaugeinvariantperturbationsfttg}

\begin{widetext}
\begin{align}
&{\mathbb{E}_{00}} \equiv F_T \left[ \frac{2 k^2 \gdep{\Psi}}{a^2} + 6 H \left( \dot{\gdep{\Psi}} - H \gdep{\Phi} \right) \right]
+ \delta F_{T_G} \left( \frac{T_G}{2} - \frac{4  k^2 H^2}{a^2} \right)
+ \dot{F_{T_G}} \left[12 H^2 \left( 4 H \gdep{\Phi} - 3 \dot{\gdep{\Psi}} \right) - \frac{8 k^2 \gdep{\beta} H^2 }{a} \right] \nonumber \\
& \quad - 12 H^3 \dot{\delta F_{T_G}} + 6 H^2 \delta F_T = -\kappa^2 \gdep{\delta\rho}\label{eq_35}\\[1.5ex]
& \int_{\hat{i}}{\mathbb{E}_{(0\,\hat{i})}} \equiv 4 H \dot{F_{T_G}} \left( \dot{\gdep{\Psi}} - H (2 \gdep{\Phi} + \gdep{\Psi}) \right)
+ F_T \left( 2 H \gdep{\Phi} - 2 \dot{\gdep{\Psi}} \right)
- \gdep{\Psi} \dot{F_T} + 4 H^2 \dot{\delta F_{T_G}} + 4 \dot{H} H \delta F_{T_G} - H \delta F_T=a \kappa^2 \gdep{U} (p + \rho)\label{eq_36}\\[1.5ex]
& \int_{\hat{i}}{\mathbb{E}_{[0\,\hat{i}]}} \equiv
 4 H \dot{F_{T_G}} \left( H (\gdep{\Phi} - \gdep{\Psi}) - \dot{\gdep{\Psi}} \right)
- \gdep{\Psi} \dot{F_T}
+ \frac{T_G \delta F_{T_G}}{6 H}
+ H \delta F_T=0\label{eq_37}\\[1.5ex]
& \int_{\hat{i}\hat{j}}\underset{{\hat{i}\neq\hat{j}}}{\mathbb{E}_{{\hat{i}\hat{j}}}}\equiv
 4 \gdep{\beta} H \ddot{F_{T_G}}
+ \dot{F_{T_G}} \left( \frac{4 H \gdep{\Psi}}{a} + \frac{\gdep{\beta} T_G}{6 H^2} + 4 \dot{\gdep{\beta}} H \right)
- \frac{F_T (\gdep{\Phi} + \gdep{\Psi})}{a} - \gdep{\beta} \dot{F_T}=0\label{eq_38}\\[1.5ex]
& {\mathbb{E}^{\hat{i}}\,_{\hat{i}}} \equiv \dot{F_{T_G}} \Biggl[
    -8 H \ddot{\gdep{\Psi}}
    - \frac{k^2 \left( a \gdep{\beta} T_G + 24 a \dot{\gdep{\beta}} H^3 + 24 H^3 \gdep{\Psi} \right)}{9 a^2 H^2}
    + \frac{4 \gdep{\Phi} T_G}{3 H}
    + 12 H^2 \dot{\gdep{\Phi}}
    - 8 (3 H^2 + \dot{H}) \dot{\gdep{\Psi}}
\Biggr]+ \frac{1}{2} T_G \delta F_{T_G} \nonumber \\
& + F_T \Biggl[
    2 \ddot{\gdep{\Psi}}
    + \frac{2 k^2 (\gdep{\Phi} + \gdep{\Psi})}{3 a^2}
    - 6 H^2 \gdep{\Phi}
    - 2 H \dot{\gdep{\Phi}}
    - 4 \dot{H} \gdep{\Phi}
    + 6 H \dot{\gdep{\Psi}}
\Biggr]+ \dot{F_T} \left(
    \frac{2 \gdep{\beta} k^2}{3 a}
    - 4 H \gdep{\Phi}
    + 2 \dot{\gdep{\Psi}}
\right)
- \frac{T_G \dot{\delta F_{T_G}}}{3 H} \nonumber \\
& - \frac{8 \gdep{\beta} H k^2 \ddot{F_{T_G}}}{3 a}
+ 2 \Biggl[
    4 H (2 H \gdep{\Phi} - \dot{\gdep{\Psi}}) \ddot{F_{T_G}}
    - 2 H^2 \ddot{\delta F_{T_G}}
    + 3 H^2 \delta F_T
    + H \dot{\delta F_T}
    + \dot{H} \delta F_T
\Biggr]= \kappa^2\gdep{\delta P}\label{eq_39}
\end{align}

\end{widetext}

 See Appendix \ref{App_A} for the expanded form of the equations. These equations are written in Fourier space, where $\delta F_T = F_{TT}\, \delta T + F_{TT_G}\, \delta T_G$ and $\delta F_{T_G} = F_{TT_G}\, \delta T + F_{T_GT_G}\, \delta T_G$, with the background teleparallel Gauss-Bonnet invariant given by $T_G = 24 H^2 (\dot{H} + H^2)$ and the perturbed scalars given by 
\begin{align}
\delta T &= \frac{ k^2}{a}4 H \gdep{\beta} + 12 H \left(  \dot{\gdep{\Psi}} -  H \gdep{\Phi} \right)\label{eq_40},\\[1.0ex]
\delta T_G &= -24 \dot{\gdep{\Phi}} H^3 + 24 \ddot{\gdep{\Psi}} H^2 - 4 \gdep{\Phi} T_G + \frac{2 \dot{\gdep{\Psi}} (24 H^4 + T_G)}{H}\nn\\ &+ \frac{2 k^2}{3 a^2 H} (12 \gdep{\Phi} H^3 + a \gdep{\beta} T_G)\label{eq_41}.
\end{align}
 All the components of the antisymmetric part \eqref{eq_24} except for the mixed components \eqref{eq_37} vanish. This is essential to constrain the Lorentz scalar $\gdep{\beta}$, which otherwise in GR does not appear. Scalar perturbations are coupled to the perturbations of the energy--momentum tensor components, and therefore, the equations alone are insufficient to fully determine the impact of cosmological perturbations on observable quantities. In the next section, we will analyse the matter perturbation equations in order to investigate the role of $F(T, T_G)$ gravity in the growth of cosmic structures.

\section{M\'esz\'aros  Equation in the Sub-Hubble era}\label{sec:meszaros}
In what follows, we examine the growth of inhomogeneous structures during the matter-dominated epoch, i.e., the period when $p=\gdep{\delta P}=0$. This is carried out by analyzing sub-horizon modes $(k \gg aH)$ and deriving the $F(T, T_G)$ counterpart of the Mészáros equation. 

Since the matter Lagrangian is invariant under general coordinate transformations, the energy--momentum tensor must be conserved with respect to the Levi-Civita connection. Consequently, one obtains two equations that have precisely the same form as their corresponding expressions in GR. 
\begin{align}
- \frac{k^2 \rho \gdep{U}}{a} + \dot{\gdep{\delta \rho}} + 3H \gdep{\delta \rho} + 3 \rho \dot{\gdep{\Psi}}=0,\,\label{eq_42}\\[1ex]
\rho \left( 4 a H \gdep{U} + a \dot{\gdep{U}} + \gdep{\Phi} \right) + a \dot{\rho} \gdep{U}=0.\label{eq_43}
\end{align}

Defining the variable $\gdep{V}=a \gdep{U}$ and using the background continuity equation \eqref{eq_B1}, the above system is transformed to 
\begin{align}
 \dot{\gdep{\delta\rho}} + 3 \gdep{\delta\rho} H - \frac{k^2 }{a^2}\rho \gdep{V}+ 3 \rho \dot{\gdep{\Psi}}&=0,\label{eq_44}\\[1ex]
 \dot{\gdep{V}} + \gdep{\Phi}&=0.\label{eq_45}
\end{align}
In terms of the gauge-invariant fractional matter perturbation $\underaccent{\mathsf{NG}}{{\hat{\delta}_m}}=\displaystyle\frac{\gdep{\delta\rho}}{\rho}+3H\gdep{V}$, the system reduces to 

\begin{align}
{\underaccent{\mathsf{NG}}{{\ddot{\hat{\delta}}_m}}}  + 2 {\underaccent{\mathsf{NG}}{{\dot{\hat{\delta}}_m}}} H  - 3 \gdep{V} \ddot{H} + 3 \ddot{\gdep{\Psi}} + \frac{k^2 }{a^2}\gdep{\Phi} + 6 H^2 \gdep{\Phi}\nonumber\\+ 6 \dot{H} (\gdep{\Phi} - H \gdep{V}) + 3 H \dot{\gdep{\Phi}} + 6 H \dot{\gdep{\Psi}}=0.\label{eq_46}
\end{align}

Where Eq. \eqref{eq_45} has been used repeatedly to get rid of the time derivatives of $\gdep{V}$. For a complete simulation of the evolution of the structure, \eqref{eq_46} should evolve with the evolution equations of the perturbations $\gdep{\Phi}$ and $\gdep{\Psi}$, when the equations have been utilized to constrain the perturbation $\gdep{\beta}$. However, for our purpose, we shall only analyze the sub-horizon modes. 

Under the sub-horizon approximation (SHA), we assume that the relevant modes have physical momenta $k/a$ that are much larger than the Hubble expansion rate $H$, while remaining well below the cutoff scale of the underlying theory. For modes deep in the horizon, a quasi-static (QSA) evolution assumption is also made, which amounts to ignoring the time derivatives of the perturbations. Under these assumptions, Eq. \eqref{eq_46} reduces to
\begin{align}
{\underaccent{\mathsf{NG}}{{\ddot{\hat{\delta}}_m}}}  + 2 H{\underaccent{\mathsf{NG}}{{\dot{\hat{\delta}}_m}}}  + \frac{k^2 }{a^2}\gdep{\Phi} =0.\label{eq_47}
\end{align}

Essentially, the terms on the right-hand side of Eq.\eqref{eq_46} contribute at order $H^2$, and are therefore negligible compared to the $(k^2/a^2)$ term for modes that lie well inside the Hubble radius. Eq. \eqref{eq_47} can then be used to compare with the standard matter perturbation equation 
\begin{equation}
    {\underaccent{\mathsf{NG}}{{\ddot{\hat{\delta}}_m}}}  + 2H {\underaccent{\mathsf{NG}}{{\dot{\hat{\delta}}_m}}} -\frac{\kappa^2}{2}\rho\frac{G_{\rm eff}}{G}{\underaccent{\mathsf{NG}}{{{\hat{\delta}}_m}}}\simeq 0\,,\label{eq_48}
\end{equation}

Where $G_{\rm eff}$ is called the effective gravitational constant. Now we need to find relation between ${\underaccent{\mathsf{NG}}{{{\hat{\delta}}_m}}}$ and $\displaystyle\frac{k^2 }{a^2}\gdep{\Phi}$ using the field equations. We adopt the approach of \cite{DeFelice:2016proca,deLaCruzDombriz:2008fr} and, specifically, that of \cite{OrjuelaQuintana:2023qsasha} to perform SHA and QSA. These approximations rely on two primary assumptions: that the gravitational potentials are approximately time-independent and that $k \gg aH$. These conditions allow us to neglect time derivatives of the potentials and discard terms proportional to $H$ multiplied by the perturbation. To determine the domain in which these approximations remain valid, it is necessary to monitor the contribution of each time-derivative term appearing in the perturbation equations. For this purpose, we introduce the dimensionless parameters
\begin{equation}
\label{eq_49}
\varepsilon \equiv \frac{a H}{k}, \quad 
\delta \equiv \frac{\dot{\varepsilon}}{\varepsilon H}, \quad 
\xi \equiv \frac{\ddot{\varepsilon}}{\varepsilon H^2}, \quad  
\chi \equiv \frac{\dddot{\varepsilon}}{\varepsilon H^3},
\end{equation}
and
\begin{subequations}
\label{eq_50}
\begin{align}
\varepsilon_\Phi \equiv \frac{\dot{\gdep{\Phi}}}{H\gdep{\Phi} }, \qquad 
\varepsilon_\Psi \equiv \frac{\dot{\gdep{\Psi}}}{H\gdep{\Psi} }, \qquad 
\varepsilon_\beta \equiv \frac{\dot{\gdep{\beta}}}{H\gdep{\beta} },\\
\chi_\Phi \equiv \frac{\dot{\varepsilon}_\Phi}{H\varepsilon_\Phi }, \qquad 
\chi_\Psi \equiv \frac{\dot{\varepsilon}_\Psi}{H\varepsilon_\Psi },\qquad
\chi_\beta \equiv \frac{\dot{\varepsilon}_\beta}{H\varepsilon_\beta }.
\end{align}
\end{subequations}
Accordingly, the condition $k \gg aH$ corresponds to $\varepsilon \ll 1$, while the assumption $\dot{\gdep{\Phi}} \approx 0$ translates into $\varepsilon_\Phi \ll 1$. We denote the quantities defined in Eqs. \eqref{eq_49} and \eqref{eq_50} as the ``SHA parameters'' and ``QSA parameters'', respectively. The QSA parameters are motivated by the slow-roll parameters commonly employed in the background analysis of inflationary models. Moreover, aside from the parameter $\varepsilon$, the remaining SHA parameters are not required to be small. As an illustration, during the matter-dominated era, the Hubble rate evolves as $H \propto \sqrt{\Omega_{m0}} a^{-3/2}$, which leads to
\begin{equation}
\label{eq_51}
\delta = -\frac{1}{2}, \quad \xi = 1, \quad \chi = -\frac{7}{2}, \qquad \text{(MD)}
\end{equation}
where $\Omega_{m0}$ denotes the present-day matter density parameter. Now, to apply this parametrization, Eqs. \eqref{eq_35}-\eqref{eq_39} need to be expanded to the contributions of $\delta T$ and $\delta T_G$, and the time derivatives must also be expanded in order to completely track all the terms proportional to $H$ or its derivatives. We keep only the leading order terms in $\varepsilon$ to account for the modes deep in the Hubble well. Also note that the QSA parameters are very small, so their coupling with the SHA parameter $\varepsilon$ is considered higher-order and thus neglected. Also, without loss of generality, any further derivative of $\varepsilon_\Phi$ is non-linear in QSA parameters, so is neglected without defining any new parameter for derivatives of $\chi_\Phi$. See \cite{Shivamgit2026} for the full expression in terms of these parameters and for the complete calculation of the $G_{\rm eff}$. Ref.~\cite{OrjuelaQuintana:2023qsasha} established a safety region $0.067\,h\,\mathrm{Mpc}^{-1}\lesssim k \lesssim
0.2\,h\,\mathrm{Mpc}^{-1}$ for scale-dependent evolution of perturbations from matter domination to the present, and showed that second-order SHA corrections have negligible impact on linear cosmological observables within this region.  The modes relevant to the growth factor computed here lie well within this safety region, and the QSA parameters $\varepsilon_\Phi$, $\varepsilon_\Psi$  and $\varepsilon_\beta$ are always suppressed by an additional factor of $\varepsilon^n\,(n\geq2$), making any QSA-breaking contribution at least third-order and negligible. There are four variables, $\gdep{\Phi},\gdep{\Psi} , \gdep{\beta} \,\text{and}\, {\underaccent{\mathsf{NG}} {{{\hat{\delta}}_m}}}$ and four equations, $\mathbb{E}_{00},\mathbb{E}_{(0\,\hat{i})},\mathbb{E}_{(\hat{i}\,\hat{j})},\,\mathbb{E}^{\hat{i}}\,_{\hat{i}}\, \text{and a constraint equation} \,\,\mathbb{E}_{[0\,\hat{i}]} $. We choose the first three equations and the constraint to solve for the variables completely. However, we checked that the fourth equation was also satisfied. Moreover, we have provided the complete expansions of all the equations in Appendix \ref{App_A}. We start by multiplying the time-space component by $3H$ and adding it to the time-time component, and then substituting for $\gdep{\beta}$ from the antisymmetric part of the equations. At leading order, this looks like
\begin{equation}\label{eq_52}
 \kappa^2\left(\gdep{\delta\rho}  + \frac{3 \varepsilon k \rho \gdep{V}}{a}\right) +\frac{2 k^2 \gdep{\Psi} F_T}{a^2} \simeq 0.
\end{equation}
In terms of  ${\underaccent{\mathsf{NG}}{{{\hat{\delta}}_m}}}$, after substituting $\varepsilon$ back, we get
\begin{equation}\label{eq_53}
 \kappa^2 \rho{\underaccent{\mathsf{NG}}{{{\hat{\delta}}_m}}}  + \frac{2 k^2 }{a^2}\gdep{\Psi} F_T \simeq 0.
\end{equation}

The anisotropic part, after substitution of $\gdep{\beta}$, at leading order in $\varepsilon$, results in 
\begin{equation}\label{eq_54}
    \gdep{\Phi}+\gdep{\Psi}+ O[\varepsilon^3]\simeq0.
\end{equation}
which is true for $F(T)$ gravity as well \cite{Zheng_2011}. This means, \eqref{eq_53} is transformed to 
\begin{equation}\label{eq_55}
 \kappa^2 \rho{\underaccent{\mathsf{NG}}{{{\hat{\delta}}_m}}}  - \frac{2 k^2 }{a^2}\gdep{\Phi} F_T \simeq 0.
\end{equation}
Which can be directly used in \eqref{eq_47} to get the $G_{\rm eff}$. But before doing so, we reproduce Eq. \eqref{eq_47} using the parametrization used to prove its robustness. In terms of SHA-QSA parameters, \eqref{eq_46} is given by 

\begin{align}
&{\underaccent{\mathsf{NG}}{{\ddot{\hat{\delta}}_m}}}+ \frac{2 k \varepsilon {\underaccent{\mathsf{NG}}{{\dot{\hat{\delta}}_m}}}}{a}  + \left( \frac{k^2}{a^2} + \frac{6 k^2 \delta \varepsilon^2}{a^2} + \frac{3 k^2 \varepsilon^2 \varepsilon_\Phi}{a^2} \right) \gdep{\Phi} \nonumber \\
& + \left( \frac{3 k^2 \varepsilon^2 \varepsilon_\Psi}{a^2} + \frac{3 k^2 \delta \varepsilon^2 \varepsilon_\Psi}{a^2} + \frac{3 k^2 \varepsilon^2 \varepsilon_\Psi^2}{a^2} + \frac{3 k^2 \varepsilon^2 \varepsilon_\Psi \chi_\Psi}{a^2} \right) \gdep{\Psi} \nonumber \\ 
&\qquad+\left( \frac{3 k^3 \delta \varepsilon^3}{a^3} - \frac{3 k^3 \varepsilon^3 \xi}{a^3} \right) \gdep{V}
 =0.\label{eq_56}
\end{align}

At leading order, this becomes

\begin{align}
{\underaccent{\mathsf{NG}}{{\ddot{\hat{\delta}}_m}}}+ \frac{2 k \varepsilon {\underaccent{\mathsf{NG}}{{\dot{\hat{\delta}}_m}}}}{a} + \frac{k^2 \gdep{\Phi}}{a^2} \simeq 0.\label{eq_57}
\end{align}

which after substituting $\varepsilon$ is exactly \eqref{eq_47}. Eliminating $\gdep{\Phi}$ from \eqref{eq_47}  and \eqref{eq_55}, we get  

\begin{equation}
    {\underaccent{\mathsf{NG}}{{\ddot{\hat{\delta}}_m}}}  + 2H {\underaccent{\mathsf{NG}}{{\dot{\hat{\delta}}_m}}} +\frac{\kappa^2\rho}{2F_T} {\underaccent{\mathsf{NG}}{{{\hat{\delta}}_m}}}\simeq 0\,.\label{eq_58}
\end{equation}

Comparing with \eqref{eq_48}, we get 
\begin{equation}
    \displaystyle G_{\rm eff}=\frac{G}{-F_T},\label{eq_59}
\end{equation}

which is of the same form as $F(T)$ gravity \cite{Zheng_2011,Chen:2010va}. 
The effects of the Gauss–Bonnet term enter through the $T_G$-dependence of $F_T$ and, via the background equations \eqref{eq_30}–\eqref{eq_31}, through the corresponding solution for the Hubble function that appears in the drag term. Furthermore, this puts forward an interesting class of theories of the form $-T+f(T_G)$, where the effective gravitational constant would be equal to $G$ as in GR, and the Gauss-Bonnet term acts as a dark energy component.  In the background, this theory is dynamically equivalent to $R+f(G)$ gravity, but at the perturbative level, the latter faces many problems, e.g., the non-luminal speed of tensor waves, which does not occur in this theory by construction \cite{Mishra:2025}. It is therefore compelling to study such a class of models in the context of the growth of structure, which we do in the next section.

\section{Growth Factor and Numerical Analysis} \label{sec:num_results}
The growth factor is defined as the ratio of the fractional matter perturbation at scale factor to its value at a reference initial scale factor $a_i$. 
\begin{equation}\label{eq_60}
\mathcal{D}(a) = \frac{{\underaccent{\mathsf{NG}}{{{\hat{\delta}}_m}}}(a)}{{\underaccent{\mathsf{NG}}{{{\hat{\delta}}_m}}}(a_i)} \, .
\end{equation}
In terms of the scale factor, the M\'esz\'aros equation \eqref{eq_58} is given by
\begin{align} \label{eq_61}
 {\underaccent{\mathsf{NG}}{{{\hat{\delta}}_m}}}''(a) +  \left( \frac{h'(a)}{h(a)} + \frac{3}{a} \right){\underaccent{\mathsf{NG}}{{{\hat{\delta}}_m}}}'(a)+\frac{3  \Omega_{m_0}{\underaccent{\mathsf{NG}}{{{\hat{\delta}}_m}}}(a)}{2 a^5 h(a)^2 F_T(a)}=0.
\end{align}

Where $h(a)=\frac{H(a)} {H_0}$ is the normalized Hubble parameter and dashes represent derivative with respect to the scale factor $a$. Furthermore, $\rho$ has been substituted using $\displaystyle \rho(t) = \frac{3 H_0^2 \Omega_{m_0}}{\kappa^2}\, a^{-3}$, where quantities with a subscript $0$ denote those at present time, where $a=1$. In the following, we analyze the evolution of $\mathcal{D}$ w.r.t. $a$. The initial condition for the ODE \eqref{eq_60} is given by ${\underaccent{\mathsf{NG}}{{{\hat{\delta}}_m}}}(a_i)=0.1$ and ${\underaccent{\mathsf{NG}}{{{\hat{\delta}}_m}}}'(a_i)=1$ in order to ensure matter dominated era, where $a_i=0.1$. The models analyzed in the subsequent section do not admit closed-form solutions; consequently, a numerical approach is adopted, requiring the computation of the function $h(a)$ for each model. So the Mészáros equation is made to evolve simultaneously with the background equations. Another very important dynamical parameter is the growth index $\gamma(a)$, which is defined by 
\begin{align} \label{eq_62}
    g(a) = \frac{d \ln {\underaccent{\mathsf{NG}}{{{\hat{\delta}}_m}}}(a)}{d \ln a} \simeq \Omega_{m}(a)^{\gamma(a)}\,,    
\end{align}
where $g(a)$ denotes the cosmic growth rate, quantifying the suppression of the linear growth factor with the expansion of the Universe. Herein, the growth factor can be re-expressed as
\begin{align}\label{eq_63}
    \mathcal{D}(a) = \exp \left( \int_{1}^{a} \frac{\Omega_{m}(\tilde{a})^{\gamma(\tilde{a})}}{\tilde{a}} d\tilde{a} \right)\,.
\end{align}

We can get the growth index $\gamma$ at the current time for each model separately. This is done by numerically differentiating the interpolating function obtained for ${\underaccent{\mathsf{NG}}{{{\hat{\delta}}_m}}}$, taking the Logarithm on both sides, and then interpolating the value at $a=1$. We chose the following models for our analysis.

\begin{align}
 \displaystyle  \textbf{A.} \quad&F(T,T_G) = -T + c_{1}\!\left(\frac{T_G}{T_{G0}}\right)^{p_1} \label{eq_64}
\\
    \textbf{ B.}  \quad&F(T,T_G) =-T\,+ c_{21}\sqrt{T^2+c_{22}T_G} \label{eq_65}\\
    \displaystyle   \textbf{C.} \quad&F(T,T_G)=-T-c_3 e^{-p_3\left(\frac{T_G}{T_{G0}}\right)} \label{eq_66}
\end{align}
The models \textbf{A} and \textbf{C} belong to the class $-T+f(T_G)$, with $f$ a power law and an exponential, respectively. We have taken model\textbf{ B} to ensure the generality of $F(T, T_G)$ and not to constrain the model space. The model parameters, i.e., the coefficients $c_1,c_{21},c_{22},c_3$ and the exponents $p_1,p_2$  are chosen using the Friedmann equation at the current time as a constraint. We also ensured that the parameters take only those values for which the stability condition $-F_T+4H\dot{F_{T_G}}>0$ was satisfied \cite{Mishra:2025}. In what follows, we compare the results with the standard cosmological model, $\Lambda$CDM.  The growth index value for  $\Lambda$CDM is $0.55$. In all analyses of the test models below, the assumptions $\Omega_{m0}=0.30$ and $H_0=70$ have been made.
\subsection{\texorpdfstring{$\displaystyle F(T,T_G) = -T + c_{1}\!\left(\frac{T_G}{T_{G0}}\right)^{p_1}
$}{po}}\label{sec:power} 
This is the simplest model one can think of \cite{Kadam:2024tgb}. A normalization factor has been introduced to make the parameter $p_1$ dimensionless, and $c_1$ has the dimension of $T$. The $\Lambda$CDM limit is obtained by substituting $p_1=0$. The parameter $c_1$ can be fixed at an arbitrary epoch. In particular, evaluating the first Friedmann Eq. \eqref{eq_30} at the present cosmic time $t_0$, defined by $a(t_0)=1$ and $H(t_0)=H_0$, yields 
\begin{equation} \label{eq_68}
c_1 = \frac{2 \cdot 3^{1-p_1} 25^{p_1} 107^{2-p_1} ({1-\Omega_{m0}} ) H_0^{2(1-2 p_1)} T_{G0}^{p_1}}{(p_1-1) (1302 p_1 + 11449)}\,.
\end{equation}

\begin{figure}[H]
    \centering
    \includegraphics[width=1\linewidth]{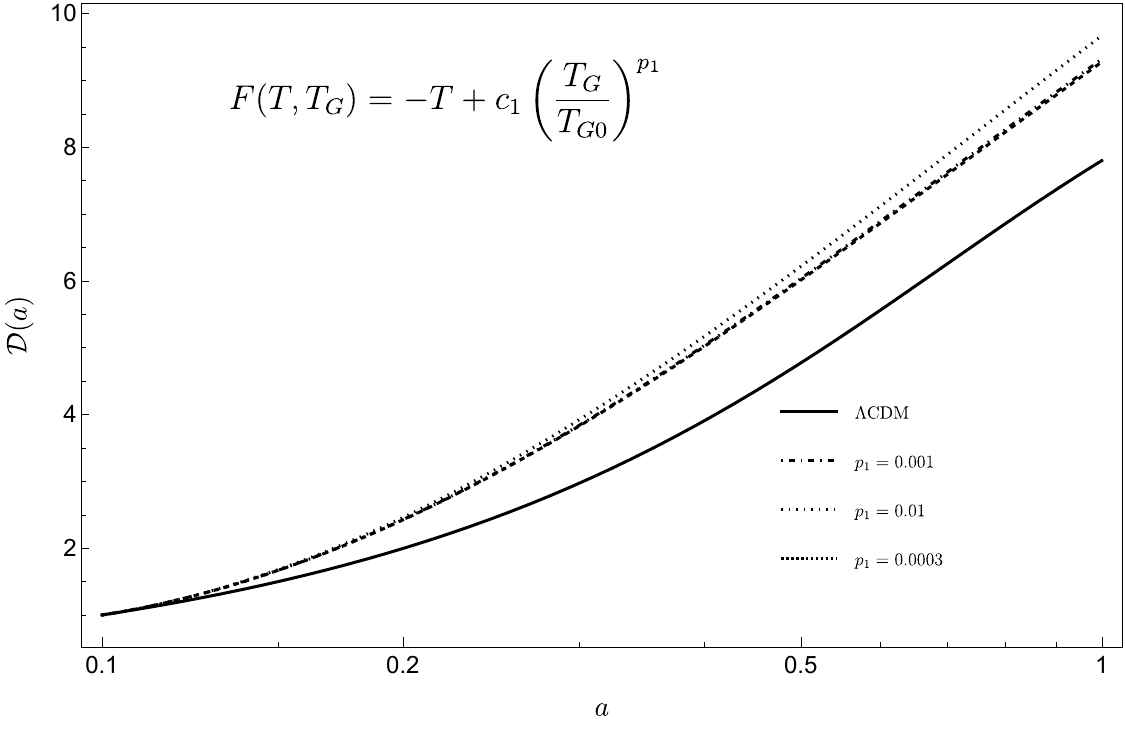}
    \caption{Growth factor evolution for model \textbf{A}. The bold line represents the growth factor for $\Lambda$CDM. The dotted-dashed, dashed, and thick dashed lines represent the growth factor for model \textbf{A} at parameter values $p_1=0.001$, $p_1=0.01$, and $p_1=0.0003$, respectively. We see that the model deviates from  $\Lambda$CDM for all values of $p_1$.} \label{fig:1}
\end{figure}

Where derivatives of the Hubble parameter are regarded as $H'(a)=H_0\,h'(a)$ and $h'(a=1)$ is obtained via using standard cosmographic relations. We use this in \eqref{eq_30} and solve it together numerically with \eqref{eq_61}. The resultant growth factor is plotted in Fig. \ref{fig:1} for three parameter values  $p_1=0.001$, $p_1=0.01$, and $p_1=0.0003$.

As illustrated in Fig. \ref{fig:1}, the power-law model departs from the $\Lambda$CDM prediction relatively early, shortly after the matter-dominated epoch. Moreover, the model continues to exhibit an enhanced growth rate up to the present time. Additionally, increasing the parameter value amplifies the growth behavior, whereas decreasing it leads to only marginal deviations from the baseline evolution.  The growth index for all three parameters listed above is 0.57 (to two decimal places). 

\subsection{\texorpdfstring{$\displaystyle F(T,T_G) =-T\,+ c_{21}\sqrt{T^2+c_{22}T_G}$}{sqrt}}\label{sec:sqrt}

The above model is among the most studied \cite{Kofinas:2014aka,Lohakare:2023tgb,Lohakare:2026tgb,Asimakis:2022tgb,Balhara:2024tgb}. The reason is that $T_G$ contains quartic torsion contributions and is therefore of the same effective order as $T^2$. As a result, $T$ and the combination $\sqrt{T^2+c\, T_G}$ naturally appear at the same order and must be included at the same effective order in the modified action. $c_{21}$ and $c_{22}$ are dimensionless coupling parameters that are expected to primarily affect late-time cosmological dynamics, while still allowing for non-trivial phenomenology within $F(T, T_G)$ cosmology. And since they are dimensionless, no new mass scale enters at late times either. Although the model does not reduce exactly to $\Lambda$CDM for finite parameter values, $c_{22}=0$ provides a rescaled TEGR action. The parameter $c_{21}$ is determined by the Friedmann equation \eqref{eq_30} at current time as  
\begin{equation} \label{eq_69}
 c_{21} =  \frac{\sqrt{3} \left(107 c_{22} + 300\right)^{3/2} \left(1-{\Omega_{m0}} \right)}{5 \left(3-c_{22}\right) \left(121 c_{22} + 600\right)}.
\end{equation}

\begin{figure}[H]
    \centering
    \includegraphics[width=1\linewidth]{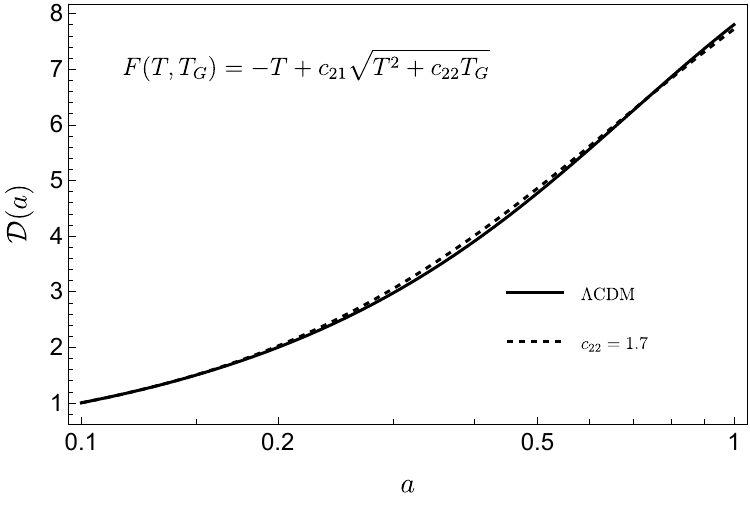}
    \caption{Growth factor evolution for model \textbf{B}. The solid curve corresponds to the $\Lambda$CDM growth factor, while the dashed-dotted curve denotes the growth factor for model \textbf{B} evaluated at $c_{22}=1.7$. The results indicate that the model closely tracks $\Lambda$CDM.
} \label{fig:2}
\end{figure}

Substituting this in \eqref{eq_30} and solving it together with \eqref{eq_61} results in Fig. \ref{fig:2}, where we have considered $c_{22}=1.7$. Note that no factors of $H_0$ appear in the constraint for $c_{21}$.

For this model, the growth index evaluated at the present epoch for the parameter value $c_{22}=1.7$ is $\gamma \approx 0.55$. As shown in Fig.~\ref{fig:2}, the model closely follows the $\Lambda$CDM behavior at both early and late times, with negligible deviations at intermediate epochs. As $a \to 1$, the growth exhibits a slight suppression relative to $\Lambda$CDM. Overall, this model remains consistent with standard cosmology at both the background and linear perturbation levels.

\subsection{\texorpdfstring{$\displaystyle F(T,T_G)=-T-c_3 e^{-p_3\left(\frac{T_G}{T_{G0}}\right)}$}{expo}}
Exponential corrections to gravitational actions have been extensively investigated in the literature \cite{Linder:2009exp, Bamba:2011ft}. As a second representative example within the class of models of the form $-T+f(T_G)$, we analyze this model to assess the perturbative viability of the class. The model parameter $p_3$ is dimensionless, while $c_3$ carries the same dimension as $T$. The latter is fixed by imposing the first Friedmann equation, yielding
\begin{equation} \label{eq_70}
    c_3=\frac{68694 H_0^2 e^{p_3} \left(\Omega_{m0} - 1\right)}{1302 p_3^2 - 11449 p_3 - 11449}.
\end{equation}

\begin{figure}[H]
    \centering
    \includegraphics[width=1\linewidth]{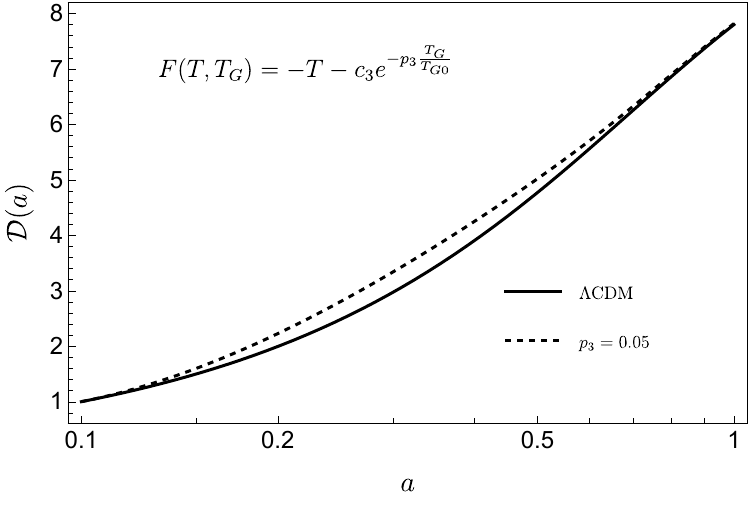}
    \caption{Growth factor evolution for model \textbf{C}. The solid curve represents the $\Lambda$CDM growth factor, while the dashed-dotted curve corresponds to the growth factor for model \textbf{C} evaluated at $p_3=0.05$. The results show that the model exhibits only minimal deviations from the $\Lambda$CDM prediction.}
 \label{fig:3}
\end{figure}

We plot the growth factor after numerically solving \eqref{eq_30} and \eqref{eq_61} with $F$ given by \eqref{eq_66} in Fig. \ref{fig:3}, where the chosen parameter value is $p_3=0.05$.

It is clear from Fig. \ref{fig:3} that early-time evolution ($a \ll 1$) is indistinguishable from $\Lambda$CDM, indicating that the modified gravity sector is subdominant during the matter era and remains consistent with standard structure formation and early-universe constraints. At late times ($a \to 1$), the model exhibits minimal enhanced growth relative to $\Lambda$CDM, consistent with a strengthened effective gravitational coupling or reduced late-time suppression of perturbation growth while exhibiting moderate deviations at intermediate epochs. Overall, the deviation remains moderate, preserving standard early-time cosmology while introducing observable late-time modifications to structure formation. The current growth index obtained for this model is $0.56$. 

\section{Conclusion} \label{sec:conclusion}
We have investigated the evolution of matter density perturbations in teleparallel gravity generalized by a Gauss-Bonnet term. The main result of this paper is the  M\'esz\'aros  Equation \eqref{eq_58}, which is obtained for sub-horizon modes. This is done using the parametrization scheme used in \cite{DeFelice:2016proca,deLaCruzDombriz:2008fr, OrjuelaQuintana:2023qsasha} with QSA-SHA parameters given by \eqref{eq_49}-\eqref{eq_50}. We see that at leading order in the sub-horizon quasi-static regime, the Gauss-Bonnet term $T_G$ does not contribute explicitly  to the effective gravitational constant $G_{\rm eff}$, offering the possibility of a model space of the type $-T+f(T_G)$, naturally reducing to GR at early times. Additionally, we have analyzed growth in three test models \eqref{eq_64}-\eqref{eq_66}, considering both minimally coupled models mentioned above and general-coupling models, while accounting for the dimensions and stability conditions of tensor and vector perturbations of the theory.

Starting from the on-shell linearized perturbation equations \eqref{eq_35}–\eqref{eq_39} in the Newtonian (longitudinal) gauge, and employing both background and perturbed continuity equations for dust, we reduced the system to the coupled equations \eqref{eq_44}–\eqref{eq_45}, which can be recast into the second-order differential equation \eqref{eq_46}. Implementing the QSA–SHA parametrization \eqref{eq_49}–\eqref{eq_50} in the geometric perturbation sector and substituting into the second-order evolution equation for the gauge-invariant fractional matter overdensity, we derive, after algebraic manipulation, the generalized M\'esz\'aros equation \eqref{eq_58}. The resulting equation is solved numerically together with the background evolution equation \eqref{eq_30} to determine the growth factor evolution \eqref{eq_60} for three representative models. These correspond to: (A) a minimal power-law extension \eqref{eq_64}, (B) a square-root model \eqref{eq_65}, and (C) a normalized exponential extension \eqref{eq_66}. We further evaluate the present-day growth index for each case.

Each model contains two free parameters, one of which is fixed through the Friedmann constraint evaluated at the present epoch, while the second is restricted by the no-ghost stability conditions arising from the tensor and vector perturbation sectors. Our results indicate that models (B) and (C) closely track $\Lambda$CDM at both early and late times, exhibiting only minimal deviations. In contrast, the power-law model (A) shows systematic deviations throughout cosmic evolution and yields a comparatively larger growth index.

We have therefore demonstrated that teleparallel Gauss–Bonnet gravity provides a viable framework for constructing dark energy models with distinctive observational imprints. 
A detailed confrontation with growth-rate data, in particular $f\sigma_8$ measurements from redshift-space distortion surveys and $S_8$ constraints from weak-lensing surveys is an important direction for future work. The analytical framework developed here provides the necessary input for such a statistical analysis.  Additionally, the galaxy--galaxy weak-lensing methodology developed for $f(T)$ gravity in Refs.~\cite{Chen:2019ftv,Wang:2023qfm} can be naturally extended to $F(T,T_G)$ models, offering a complementary observational handle on the parameter space beyond background-level probes. Another significant extension would be to derive the fully general evolution equation for the matter overdensity without invoking sub-horizon or quasi-static approximations, and to compare the resulting dynamics with those obtained in the present analysis to assess the validity and regime of applicability of these approximation schemes.
\section*{Acknowledgments} B.M. acknowledges the support of Anusandhan National Research Foundation(ANRF) for the grant (File No: CRG/2023/000475). This article is also based on work from COST Action CA21136 Addressing observational tensions in cosmology with systematics and fundamental physics (CosmoVerse), supported by COST (European Cooperation in Science and Technology). SKM acknowledges the International Centre for Theoretical Sciences (ICTS) for the program ICTS Summer School on Gravitational-Wave Astronomy (ICTS/GWA2025/07), during which part of this work was carried out. JLS would also like to acknowledge funding from ``Xjenza Malta'' as part of the ``Technology Development Programme'' DTP-2024-014 (CosmicLearning) Project.

\appendix
\section{Expanded Field Equations}\label{App_A}
Using the relations  $\delta F_T = F_{TT}\, \delta T + F_{TT_G}\, \delta T_G$, $\delta F_{T_G} = F_{TT_G}\, \delta T + F_{T_GT_G}\, \delta T_G$,  $\dot{F_{T}}=F_{TT}\dot{T}+F_{TT_{G}}\dot{T}_{G}$,
$\dot{F_{T_{G}}}=F_{TT_{G}}\dot{T}+F_{T_{G}T_{G}}\dot{T}_{G}$ and 
$\ddot{F_{T_{G}}}=F_{TTT_{G}}\dot{T}^{2}+2F_{TT_{G}T_{G}}\dot{T}
\dot{T}_{G}+F_{T_{G}T_{G}T_{G}}\dot{T}_{G}^{\,\,2}+
F_{TT_{G}}\ddot{T}+F_{T_{G}T_{G}}\ddot{T}_{G}$ and substituting $T_G$ as $24 H^{2} (\dot{H} + H^{2})\,$ in \eqref{eq_40}-\eqref{eq_41}, we get the equations \eqref{eq_35} to \eqref{eq_39} as

\begin{widetext}
Time-time component ${(\mathbb{E}_{00})}\rightarrow$
    \begin{align}
& F_{T_G T_G} \Biggl[
    -288 H^3 \Bigl[
        H \Bigl(
            H \Bigl(
                -H \ddot{\gdep{\Phi}} - 8 \gdep{\Phi} \ddot{H} + 3 H \ddot{\gdep{\Psi}} + \dddot{\gdep{\Psi}} + 4 H^3 \gdep{\Phi}
            \Bigr)
            + \dot{\gdep{\Psi}} (5 \ddot{H} - 4 H^3)
            - 3 H^3 \dot{\gdep{\Phi}}
        \Bigr)
        - 3 H \dot{H} \Bigl(
            -\ddot{\gdep{\Psi}} + 8 H^2 \gdep{\Phi} + 2 H \dot{\gdep{\Phi}}
        \nonumber \\
& \quad   - 6 H \dot{\gdep{\Psi}}
        \Bigr)+ 6 \dot{H}^2 (\dot{\gdep{\Psi}} - 2 H \gdep{\Phi})
    \Bigr]+ \frac{96 H^3 k^2}{a^2} \Bigl[
        H \Bigl(
            -4 a \gdep{\beta} \ddot{H} - \ddot{\gdep{\Psi}} + 4 a \gdep{\beta} H^3 - 2 a \dot{\gdep{\beta}} H^2 + 7 H^2 \gdep{\Phi} - 4 H \dot{\gdep{\Psi}}
        \Bigr)
        - \dot{H} \Bigl(
            8 a \gdep{\beta} H^2 + 2 a \dot{\gdep{\beta}} H
        \nonumber \\
& \quad  - 3 H \gdep{\Phi} + 2 \dot{\gdep{\Psi}}
        \Bigr)  - 4 a \gdep{\beta} \dot{H}^2
    \Bigr] + \frac{32 H^3 k^4}{a^4} \Bigl(
        H (2 a \gdep{\beta} H + \gdep{\Phi}) + 2 a \gdep{\beta} \dot{H}
    \Bigr)\Biggr]+ F_{TTT_G} \Biggl[
    1728 H^5 \dot{H} (H \gdep{\Phi} - \dot{\gdep{\Psi}})
    - \frac{576 \gdep{\beta} H^5 \dot{H} k^2}{a}
\Biggr]
 \nonumber \\
& + F_{T_G T_G T_G} \Biggl[
    6912 H^5 (H \ddot{H} + 4 \dot{H} H^2 + 2 \dot{H}^2) \Bigl(
        H (-\ddot{\gdep{\Psi}} + 4 H^2 \gdep{\Phi} + H \dot{\gdep{\Phi}} - 4 H \dot{\gdep{\Psi}})
        + \dot{H} (4 H \gdep{\Phi} - 2 \dot{\gdep{\Psi}})
    \Bigr) \nonumber \\
& \quad - \frac{2304 H^5 k^2 (H \ddot{H} + 4 \dot{H} H^2 + 2 \dot{H}^2) (H (2 a \gdep{\beta} H + \gdep{\Phi}) + 2 a \gdep{\beta} \dot{H})}{a^2}
\Biggr] + F_{TT} \left[
    \frac{24 \gdep{\beta} H^3 k^2}{a} + 72 H^3 (\dot{\gdep{\Psi}} - H \gdep{\Phi})
\right]\nonumber \\
& + F_{TT_G T_G} \Biggl[
    3456 H^5 \Bigl(
        H \dot{H} (-\ddot{\gdep{\Psi}} + 8 H^2 \gdep{\Phi} + H \dot{\gdep{\Phi}} - 8 H \dot{\gdep{\Psi}})
        + H \ddot{H} (H \gdep{\Phi} - \dot{\gdep{\Psi}})
        + \dot{H}^2 (6 H \gdep{\Phi} - 4 \dot{\gdep{\Psi}})
    \Bigr) \nonumber \\
& \quad - \frac{1152 H^5 k^2 (a \gdep{\beta} H \ddot{H} + H \dot{H} (6 a \gdep{\beta} H + \gdep{\Phi}) + 4 a \gdep{\beta} \dot{H}^2)}{a^2}
\Biggr] + F_T \left[
    \frac{2 k^2 \gdep{\Psi}}{a^2} + 6 H (\dot{\gdep{\Psi}} - H \gdep{\Phi})
\right]  \nonumber \\
& + F_{TT_G} \Biggl[
    \frac{16 \gdep{\beta} H^3 k^4}{a^3}
    + \frac{48 H^3 k^2 (2 H (2 a \gdep{\beta} H + \gdep{\Phi}) - a \dot{\gdep{\beta}} H - \dot{\gdep{\Psi}})}{a^2}
    - 144 H^3 (5 H^2 - \dot{H}) (H \gdep{\Phi} - \dot{\gdep{\Psi}})
\Biggr]=- \kappa^2 \gdep{\delta \rho }
\end{align}

Time-Space symmetric Part $\left[{\mathbb{E}_{(0\,\hat{i})}}\,\text{after integration}\right]\rightarrow$
\begin{align}
& F_{T_G T_G} \Biggl[
    96 H^2 \Bigl(
        -H \Bigl(
            H \Bigl(
                H \ddot{\gdep{\Phi}} - 4 H \ddot{\gdep{\Psi}} - \dddot{\gdep{\Psi}}
            \Bigr)
            + \ddot{H} (6 H \gdep{\Phi} + H \gdep{\Psi} - 3 \dot{\gdep{\Psi}})
            + 4 H^3 \dot{\gdep{\Phi}}
        \Bigr)- 2 \dot{H}^2 \Bigl(
            H (8 \gdep{\Phi} + \gdep{\Psi}) - 3 \dot{\gdep{\Psi}}
        \Bigr)
 \nonumber + H \dot{H} \Bigl(
            5 \ddot{\gdep{\Psi}}\\
& \quad - 28 H^2 \gdep{\Phi} - 4 H^2 \gdep{\Psi} - 8 H \dot{\gdep{\Phi}} + 20 H \dot{\gdep{\Psi}}
        \Bigr)     
    \Bigr) - \frac{32 H^2 k^2}{a^2} \Bigl(
        H \Bigl(
            -2 a \gdep{\beta} \ddot{H} + 2 a \gdep{\beta} H^3 - 2 a \dot{\gdep{\beta}} H^2 + 2 H^2 \gdep{\Phi} - H \dot{\gdep{\Phi}}
        \Bigr) - H \dot{H} (2 a \dot{\gdep{\beta}} \nonumber \\
& \quad + 6 a \gdep{\beta} H + 3 \gdep{\Phi})
        - 4 a \gdep{\beta} \dot{H}^2
    \Bigr)
\Biggr]+ F_{TTT_G} \Biggl[
    \frac{192 \gdep{\beta} \dot{H} H^4 k^2}{a}
    + 576 \dot{H} H^4 (\dot{\gdep{\Psi}} - H \gdep{\Phi})
\Biggr] + F_{TT_G} \Biggl[ 4 H^4 \gdep{\Phi} - H \gdep{\Psi} \ddot{H} - 48 H \dot{H}^2 \gdep{\Psi} \nonumber \\
&
   \quad  24 H \Bigl(
        H^2 (\ddot{\gdep{\Psi}} - H (\dot{\gdep{\Phi}} + 4 \dot{\gdep{\Psi}}))
        - 6 \dot{H} H^2 (\gdep{\Phi} + \gdep{\Psi})
        + 4 \dot{H} H \dot{\gdep{\Psi}}
           \Bigr)
    - \frac{8 H^2 k^2 (H (-2 a \dot{\gdep{\beta}} + 4 a \gdep{\beta} H + \gdep{\Phi}) - 2 a \gdep{\beta} \dot{H})}{a^2}
\Biggr] \nonumber \\
& + F_{TT} \Biggl[
    12 H \Bigl(
        H (H \gdep{\Phi} - \dot{\gdep{\Psi}}) - \dot{H} \gdep{\Psi}
    \Bigr)
    - \frac{4 \gdep{\beta} H^2 k^2}{a}
\Biggr]  + F_{TT_G T_G}\Biggl[
    \frac{384 H^4 k^2 (a \gdep{\beta} H \ddot{H} + H \dot{H} (6 a \gdep{\beta} H + \gdep{\Phi}) + 4 a \gdep{\beta} \dot{H}^2)}{a^2} \nonumber \\
& \quad - 1152 H^4 \Bigl(
        H \dot{H} (-\ddot{\gdep{\Psi}} + 8 H^2 \gdep{\Phi} + H \dot{\gdep{\Phi}} - 8 H \dot{\gdep{\Psi}})
        + H \ddot{H} (H \gdep{\Phi} - \dot{\gdep{\Psi}})
        + \dot{H}^2 (6 H \gdep{\Phi} - 4 \dot{\gdep{\Psi}})
    \Bigr)
\Biggr]+ F_T \Bigl(
    2 H \gdep{\Phi} - 2 \dot{\gdep{\Psi}}
\Bigr) \nonumber \\
&+ F_{T_G T_G T_G} \Biggl[ - 2304 H^4 (H \ddot{H} + 4 \dot{H} H^2 + 2 \dot{H}^2) \Bigl(
        H (-\ddot{\gdep{\Psi}} + 4 H^2 \gdep{\Phi} + H \dot{\gdep{\Phi}} - 4 H \dot{\gdep{\Psi}})
        + \dot{H} (4 H \gdep{\Phi} - 2 \dot{\gdep{\Psi}})
    \Bigr)
     \nonumber \\
& \quad\frac{768 H^4 k^2 (H \ddot{H} + 4 \dot{H} H^2 + 2 \dot{H}^2) (H (2 a \gdep{\beta} H + \gdep{\Phi}) + 2 a \gdep{\beta} \dot{H})}{a^2}
\Biggr]= a \kappa^2 \gdep{U} (P + \rho)
\end{align}
Time-Space antisymmetric Part $\left[{\mathbb{E}_{[0\,\hat{i}]}}\,\text{after integration}\right]\rightarrow$

\begin{align}
& F_{TT_G} \Biggl[
    \frac{2 k^2 (4 a \gdep{\beta} (H^2 + \dot{H}) + H \gdep{\Phi})}{a^2}
    - \frac{6}{H} \Bigl(
        H \Bigl(
            -H \ddot{\gdep{\Psi}} + \gdep{\Psi} \ddot{H} + 6 H^3 \gdep{\Phi} + H^2 \dot{\gdep{\Phi}} - 6 H^2 \dot{\gdep{\Psi}}
        \Bigr)+ 2 H \dot{H} (2 H \gdep{\Phi} + 3 H \gdep{\Psi} - \dot{\gdep{\Psi}}) \nonumber \\
& \quad         + 2 \dot{H}^2 \gdep{\Psi}
    \Bigr)
\Biggr] + F_{T_G T_G} \Biggl[
    24 H \Bigl(
        \dot{\gdep{\Psi}} (-\ddot{H} + 4 H^3 + 2 \dot{H} H)
        + (H^2 + \dot{H}) \ddot{\gdep{\Psi}}
        + H \gdep{\Phi} \ddot{H} - H \gdep{\Psi} \ddot{H}
        - 4 H^4 \gdep{\Phi} - (H^2 + \dot{H}) H \dot{\gdep{\Phi}}  \nonumber \\
& \quad - 2 \dot{H} (2 H^2 + \dot{H}) \gdep{\Phi}
        - 2 \dot{H} (2 H^2 + \dot{H}) \gdep{\Psi}
    \Bigr)
    + \frac{8 (H^2 + \dot{H}) k^2 (H (2 a \gdep{\beta} H + \gdep{\Phi}) + 2 a \gdep{\beta} \dot{H})}{a^2}
\Biggr] \nonumber \\
& + F_{TT} \left[
    \frac{\gdep{\beta} k^2}{a} - 3 H \gdep{\Phi} - \frac{3 \dot{H} \gdep{\Psi}}{H} + 3 \dot{\gdep{\Psi}}
\right]=0
\end{align}

Space-Space anisotropic Part $\left[{\mathbb{E}_{\hat{i}\,\hat{j}}} \,\,(\hat{i}\neq\hat{j})\,\,\text{after integration}\right]\rightarrow$

\begin{align}
& 192 H F_{T_G T_G} \Biggl[
    \dot{H} \Bigl(
        7 a \gdep{\beta} H \ddot{H} + 4 H^3 (a \dot{\gdep{\beta}} + a \gdep{\beta} H + \gdep{\Psi})
    \Bigr)
    + H^2 \ddot{H} (a \dot{\gdep{\beta}} + 5 a \gdep{\beta} H + \gdep{\Psi}) + a \gdep{\beta} H^2 \dddot{H}
    + 2 H \dot{H}^2 (a \dot{\gdep{\beta}} + 9 a \gdep{\beta} H + \gdep{\Psi}) \nonumber \\
& \quad
    + 4 a \gdep{\beta} \dot{H}^3
\Biggr] + 4608 a \gdep{\beta} H^3 (H \ddot{H} + 4 \dot{H} H^2 + 2 \dot{H}^2)^2 F_{T_G T_G T_G} + 4608 a \gdep{\beta} \dot{H} H^3 (H \ddot{H} + 4 \dot{H} H^2 + 2 \dot{H}^2) F_{TT_G T_G} \nonumber \\
& + 48 H F_{TT_G} \Bigl[
    a \gdep{\beta} H \ddot{H} + 2 H \dot{H} (a \dot{\gdep{\beta}} - a \gdep{\beta} H + \gdep{\Psi}) + 2 a \gdep{\beta} \dot{H}^2
\Bigr] + 1152 a \gdep{\beta} \dot{H}^2 H^3 F_{TTT_G} - 24 a \gdep{\beta} \dot{H} H F_{TT} - 2 F_T (\gdep{\Phi} + \gdep{\Psi})=0
\end{align}
Spatial Trace Part $\left[{\mathbb{E}^{\hat{i}}\,_{\hat{i}}}\right]\rightarrow$

\begin{align}
&  \ddot{F_{T_G T_G}} H^3 \Biggl[\frac{32k^2}{a^2}\left( H (2 a H \gdep{\beta} + \gdep{\Phi}) + 2 a \gdep{\beta} \dot{H} \right) - 96 \Bigl(  \dot{H} (4 H \gdep{\Phi} - 2 \dot{\gdep{\Psi}})
    + H \Bigl(H (4 H \gdep{\Phi} + \dot{\gdep{\Phi}} - 4 \dot{\gdep{\Psi}})
        - \ddot{\gdep{\Psi}}\Bigr)\Bigr)\Biggr] + F_{T_G T_G} \Biggl[
            -63 \dot{\gdep{\Phi}} H^4\nonumber \\
& \quad + \Bigl(
                -40 \gdep{\Phi} H^3 - 12 \ddot{\gdep{\Phi}} H + 45 \ddot{\gdep{\Psi}} H - 64 \gdep{\Phi} \ddot{H} + 8 \dddot{\gdep{\Psi}}
            \Bigr) H^2 \dot{H}
            + 6 \dot{\gdep{\Psi}} (5 H^3 + 4 \ddot{H})H \dot{H}+96 H \Bigl(
        12 (\dot{\gdep{\Psi}} - 4 H \gdep{\Phi}) \dot{H}^3 + 6 H \Bigl( - 7 H \dot{\gdep{\Phi}}
     \nonumber \\
& \quad 3 (-10 \gdep{\Phi} H^2 + 5 \dot{\gdep{\Psi}} H + \ddot{\gdep{\Psi}})
                   \Bigr) \dot{H}^2+ H^2 \Bigl(
            12 \gdep{\Phi} H^5 - 6 \ddot{\gdep{\Phi}} H^3 + 5 \ddot{\gdep{\Psi}} H^3
            - 48 \gdep{\Phi} \ddot{H} H^2 - \dddot{\gdep{\Phi}} H^2 + 6 \dddot{\gdep{\Psi}} H^2  - 8 \gdep{\Phi} \dddot{H} H + \ddddot{\gdep{\Psi}} H + 8 \ddot{H} \ddot{\gdep{\Psi}}\nonumber \\
& \quad            - \dot{\gdep{\Phi}} (5 H^4 + 14 \ddot{H} H)
            + \dot{\gdep{\Psi}} (-12 H^4 + 30 \ddot{H} H + 4 \dddot{H})\Bigr)
    \Bigr)- \frac{32 H k^2}{a^2} \Bigl(
        -12 a \gdep{\beta} \dot{H}^3
        - 2 H (24 a H \gdep{\beta} + 3 \gdep{\Phi} + 2 \gdep{\Psi} + 6 a \dot{\gdep{\beta}}) \dot{H}^2 +\nonumber \\
& \quad  H \Bigl(
            H^2 (12 a H \gdep{\beta} + 13 \gdep{\Phi} - 8 \gdep{\Psi})
            - 2 (12 a \dot{\gdep{\beta}} H^2 + 3 \dot{\gdep{\Phi}} H + a \ddot{\gdep{\beta}} H + 12 a \gdep{\beta} \ddot{H})
        \Bigr) \dot{H}+ H^2 \Bigl(
            (8 a H \gdep{\beta} + 3 \gdep{\Phi}) H^3  - (2 a H \ddot{\gdep{\beta}} + \ddot{\gdep{\Phi}}) H \nonumber \\
& \quad 
           + 2 \dot{\gdep{\Phi}} H^2 - 2 (8 a H \gdep{\beta} + \gdep{\Phi} + \gdep{\Psi} + 3 a \dot{\gdep{\beta}}) \ddot{H}
            - 4 a \gdep{\beta} \dddot{H}
        \Bigr)
    \Bigr)
\Biggr] + F_T \Biggl[
    -\frac{2 (\gdep{\Phi} + \gdep{\Psi}) k^2}{3 a^2}
    + 4 \gdep{\Phi} \dot{H}
    + 2 H (3 H \gdep{\Phi} + \dot{\gdep{\Phi}} - 3 \dot{\gdep{\Psi}})
    - 2 \ddot{\gdep{\Psi}}
\Biggr]+ \nonumber \\
&   F_{TT_G T_G}\Biggl[
    \frac{384 k^2}{a^2} \Bigl(
        2 \dot{H} \Bigl(
            (-4 a \gdep{\beta} H^2 - 3 \gdep{\Phi} H + 6 a \dot{\gdep{\beta}} H + \dot{\gdep{\Phi}}) H^2- 2 H \Bigl(
            7 \dot{\gdep{\Phi}} H^3
            + (H \ddot{\gdep{\Phi}} - 7 H \ddot{\gdep{\Psi}} - \dddot{\gdep{\Psi}}) H
            + (9 H \gdep{\Phi} - 5 \dot{\gdep{\Psi}}) \ddot{H}
        \Bigr) \dot{H} \nonumber \\
& \quad + 2 \dot{H} (9 a H \gdep{\beta} + \gdep{\Phi} + 2 a \dot{\gdep{\beta}}) H + 10 a \gdep{\beta} \dot{H}^2
        \Bigr)+ 1152 \Bigl(
        (20 \dot{\gdep{\Psi}} - 44 H \gdep{\Phi}) \dot{H}^3
        + 2 H \Bigl(
            6 (\ddot{\gdep{\Psi}} - 7 H^2 \gdep{\Phi}) - 9 H \dot{\gdep{\Phi}} + 28 H \dot{\gdep{\Psi}}
        \Bigr) \dot{H}^2 \nonumber \\
& \quad + H (10 a \gdep{\beta} \dot{H} - H (2 a H \gdep{\beta} + \gdep{\Phi} - 2 a \dot{\gdep{\beta}})) \ddot{H}
    \Bigr) H^3+ H^2 \ddot{H} \Bigl(
            H (2 H \gdep{\Phi} - \dot{\gdep{\Phi}} - 2 \dot{\gdep{\Psi}}) + \ddot{\gdep{\Psi}}
        \Bigr)
    \Bigr) H^3
\Biggr] + \dot{F_{TT}} \Biggl[
    24 H^2 (H \gdep{\Phi} - \dot{\gdep{\Psi}})
    \nonumber \\
& \quad - \frac{8 H^2 k^2 \gdep{\beta}}{a}
\Biggr]+ F_{T_G T_G T_G} \Biggl[
    \frac{1536 H^3 k^2 (2 \dot{H} (2 H^2 + \dot{H}) + H \ddot{H})}{a^2} \Bigl(
        6 a \gdep{\beta} \dot{H}^2+ H (3 (4 a H \gdep{\beta} + \gdep{\Phi}) + 2 a \dot{\gdep{\beta}}) \dot{H}+ H \Bigl(
            -\gdep{\Phi} H^2 \nonumber \\
& \quad + (2 a H \dot{\gdep{\beta}} + \dot{\gdep{\Phi}}) H + 3 a \gdep{\beta} \ddot{H}
        \Bigr)
    \Bigr) - 4608 H^3 (2 \dot{H} (2 H^2 + \dot{H}) + H \ddot{H}) \Bigl(
        2 (8 H \gdep{\Phi} - 3 \dot{\gdep{\Psi}}) \dot{H}^2 + H (32 \gdep{\Phi} H^2 + 8 \dot{\gdep{\Phi}} H - 22 \dot{\gdep{\Psi}} H\nonumber \\
& \quad  - 5 \ddot{\gdep{\Psi}}) \dot{H} + H \Bigl(
            5 \dot{\gdep{\Phi}} H^3
            + (4 \gdep{\Phi} H^3 + \ddot{\gdep{\Phi}} H - 5 \ddot{\gdep{\Psi}} H + 6 \gdep{\Phi} \ddot{H} - \dddot{\gdep{\Psi}}) H  - \dot{\gdep{\Psi}} (4 H^3 + 3 \ddot{H})
        \Bigr)
    \Bigr)
\Biggr] + \ddot{F_{TT_G}} \Biggl[ 48 (\dot{\gdep{\Psi}} - H \gdep{\Phi}) H^3
 \nonumber \\
&\quad+\frac{16 k^2 \gdep{\beta} H^3}{a}\Biggr] + F_{TT} \Biggl[
    24 H \Bigl(
        \dot{H} (5 H \gdep{\Phi} - 3 \dot{\gdep{\Psi}})
        + H \Bigl(
            H (3 H \gdep{\Phi} + \dot{\gdep{\Phi}} - 3 \dot{\gdep{\Psi}}) - \ddot{\gdep{\Psi}}
        \Bigr)
    \Bigr)  - \frac{8 H k^2 (3 \gdep{\beta} \dot{H} + H (2 H \gdep{\beta} + \dot{\gdep{\beta}}))}{a}
\Biggr]+ \nonumber \\
&\quad  F_{TTT_G} \Biggl[
    576 H^3 \dot{H} \Bigl(
        (4 \dot{\gdep{\Psi}} - 6 H \gdep{\Phi}) \dot{H}
        + H \Bigl(
            H (2 H \gdep{\Phi} - \dot{\gdep{\Phi}} - 2 \dot{\gdep{\Psi}})
            + \ddot{\gdep{\Psi}}
        \Bigr)
    \Bigr)  - \frac{192 H^3 k^2 \dot{H}}{a^2} \Bigl(
        H (2 a H \gdep{\beta} + \gdep{\Phi} - 2 a \dot{\gdep{\beta}})
        - 4 a \gdep{\beta} \dot{H}
    \Bigr)
\Biggr]\nonumber \\
& \quad+ F_{TT_G} \Biggl[
    48 H^2 \Bigl(
        5 \Bigl(
            H (3 H \gdep{\Phi} + \dot{\gdep{\Phi}} - 3 \dot{\gdep{\Psi}}) - \ddot{\gdep{\Psi}}
        \Bigr) H^2 - \frac{16 H^2 k^2}{a^2} \Bigl(
        \dot{H} (20 a H \gdep{\beta} + 3 \gdep{\Phi} - 2 \gdep{\Psi} - 4 a \dot{\gdep{\beta}})
        + H \Bigl(
            H (8 a H \gdep{\beta} + \gdep{\Phi})- a \ddot{\gdep{\beta}} \nonumber \\
& \quad + 2 a H \dot{\gdep{\beta}} + \dot{\gdep{\Phi}} 
        \Bigr)
    \Bigr) - 2 \gdep{\Phi} \dot{H}^2
        + \dot{H} \Bigl(
            H (31 H \gdep{\Phi} - \dot{\gdep{\Phi}} - 21 \dot{\gdep{\Psi}})
            + \ddot{\gdep{\Psi}}
        \Bigr)
    \Bigr) \Biggr] = - \kappa^2 \gdep{\delta P}
\end{align}

We have not expanded some time derivatives of the background contributions for simplicity, as this does not affect the analysis because the trace equation is not used in the calculations.

\end{widetext}
\section{Matter Conservation}\label{App_B}

Since we are interested in the dynamics of structure in the universe, the assumption  $p=\gdep{\delta P}=0$ is employed in what follows.
\subsection{Background}
The components of matter are conserved independently, leading to the standard conservation equation for dust
\begin{equation}
    \bar{\nabla}_{\beta}\Theta_{0}\,^{\beta}:\qquad\dot{\rho}+3H\rho=0\,.\label{eq_B1}
\end{equation}
\subsection{Perturbations}
Here are first-order expansions of the components of the continuity equation: 
\begin{align}
  \delta \,\bar{\nabla}_{\beta}\Theta_{0}\,^{\beta}&:&\dot{\gdep{\delta \rho}} + 3H \gdep{\delta \rho} - \frac{k^2 \rho \gdep{U}}{a}+ 3 \rho \dot{\gdep{\Psi}}=0\,\label{eq_B2}\\[1ex]
 \delta \,\bar{\nabla}_{\beta}\Theta_{\hat{i}}\,^{\beta}&:& \partial_{\hat{i}}\left[\rho \left( 4 a H \gdep{U} + a \dot{\gdep{U}} + \gdep{\Phi} \right) + a \dot{\rho} \gdep{U}\right]=0\label{eq_B3}
\end{align}
Throughout the above expressions, the Einstein summation convention has been employed.
\bibliographystyle{utphys}
\bibliography{references}

\end{document}